\documentclass[final,1p,times]{elsarticle} 

\usepackage{graphicx}
\usepackage{amssymb} 
\usepackage{amsthm} 

\def\snn{\mbox{$\sqrt{s_{_{\rm NN}}}$}}

\def\pt{p_{\rm{T}}}

\def\psin{\Psi_{\rm n}}

\journal{Nuclear Physics A} 

\begin{document}

\begin{frontmatter} 

\title{Event shape engineering with ALICE}

\author{A.~Dobrin (for the ALICE\fnref{col1} Collaboration)}
\fntext[col1] {A list of members of the ALICE Collaboration and acknowledgements can be found at the end of this issue.}

\address{Department of Physics and Astronomy, Wayne State University, 666 W. Hancock, Detroit, Michigan 48201}

\begin{abstract}
The strong fluctuations in the initial energy density of heavy-ion 
collisions allow an efficient selection of events corresponding 
to a specific initial geometry. For such ``shape engineered events'', 
the elliptic flow coefficient, $v_2$, of unidentified 
charged particles, pions and (anti-)protons in Pb-Pb collisions at 
$\snn = 2.76$~TeV is measured by the ALICE collaboration. $v_2$ 
obtained with the event plane method at mid-rapidity, $|\eta|<0.8$, 
is reported for different collision centralities as a function of 
transverse momentum, $\pt$, out to $\pt=20$~GeV/$c$. The measured 
$v_2$ for the shape engineered events is significantly larger or smaller 
than the average which demonstrates the ability to experimentally select 
events with the desired shape of the initial spatial asymmetry.
\end{abstract}


\end{frontmatter} 


\section{Introduction}

An important observable used for the characterization of the 
properties and the evolution of the system created in a 
nucleus-nucleus collision is the anisotropic 
flow~\cite{Voloshin:2008dg}. Anisotropic flow arises due to the 
asymmetry in the initial geometry of the collision and is 
characterized by the Fourier coefficients~\cite{Voloshin:1994mz, Poskanzer:1998yz}:
\begin{equation}
 v_n(\pt,\eta) = \langle \cos[n(\phi-\psin)] \rangle,
\end{equation}
where $\pt$, $\eta$, and $\phi$ are the particle's transverse
momentum, pseudo-rapidity, and the azimuthal angle, respectively, and
$\psin$ is the $n$-th harmonic symmetry plane angle. The second 
Fourier coefficient $v_2$ is called elliptic flow. Recently, 
experimental measurements~\cite{vn_alice} confirmed the existence of 
non-zero odd harmonic coefficients due to fluctuations in the initial 
energy density distribution.

Two approaches are currently utilized to study the effect of the 
initial geometry on final observables: variation of the collision 
centrality and collisions between nuclei of different size and shape. 
A new method to select events corresponding to different initial 
system shapes based on the fluctuations in the initial geometry was 
proposed in~\cite{Schukraft:2012ah}. In this paper, 
following~\cite{Schukraft:2012ah}, we select events with elliptic 
flow values significantly larger or smaller than the average. For 
those events, we present the measurement of unidentified charged 
particle $v_2$ out to $\pt=20$~GeV/$c$, and for protons and charged 
pions~\footnote{In this analysis we do not differentiate between 
particle and antiparticle.} out to $\pt=16$~GeV/$c$.

\section{Analysis details}

The data sample recorded by ALICE during the 2010 heavy-ion run at 
the Large Hadron Collider is used for this analysis. The Time 
Projection Chamber (TPC) was used to reconstruct charged particle 
tracks and measure their momenta with full azimuthal coverage in the 
pseudo-rapidity range $|\eta|<0.8$, and particle identification via 
the specific ionization energy loss, $\mathrm{d}E/\mathrm{d}x$, in 
the transverse momentum region $\pt > 3$ 
GeV/$c$~\cite{Abelev:2012di}. Two scintillator 
arrays (VZERO) which cover the pseudo-rapidity ranges 
$-3.7<\eta<-1.7$ (VZERO-C) and $2.8<\eta<5.1$ (VZERO-A) were used for 
triggering, centrality~\cite{Aamodt:2010cz} and symmetry plane 
determination. The trigger conditions and the event selection 
criteria are identical to those described 
in~\cite{vn_alice, Aamodt:2010cz}. Approximately $1.1 \times 10^7$ 
minimum-bias Pb-Pb events with a reconstructed primary vertex within 
$\pm 10$ cm from the nominal interaction point in the beam direction 
are used for this analysis. Charged particles reconstructed in the 
TPC in $|\eta|<0.8$ and $0.2<\pt<20$ GeV/$c$ which pass the quality 
cuts described in~\cite{Aamodt:2010pa} were selected.

The event shape analysis is performed with the three subevents technique. 
The first subevent ``a'' is used for the event selection based on the 
magnitude of the so-called reduced flow vector, $q_2$~\cite{Voloshin:2008dg, Poskanzer:1998yz}:
\begin{eqnarray}
 && Q_{2,x} = \sum_i^M \cos(2 \phi_i), \quad Q_{2,y} = \sum_i^M \sin(2 \phi_i),\\
 && q_2 = Q_{2}/\sqrt{M},
\end{eqnarray}
\noindent where $M$ is the multiplicity and $\phi_i$ is the azimuthal 
angle of particle $i$. $v_2$ is measured with the event plane method 
($v_2\{\rm{EP}\}$~\cite{Voloshin:2008dg}) based on particles 
reconstructed in the second subevent ``b'' with the symmetry plane 
$\Psi_{2}$ determined from particles of the third subevent ``c''. Two 
different sets of subevents were considered: one configuration with 
``a'' and ``b'' subevents determined by two $\eta$-subevents of TPC 
tracks and ``c'' from the VZERO detector, and another configuration 
with ``a'' and ``c'' subevents from the two VZERO scintillators and 
``b'' from TPC tracks. In the latter case, the large gap in 
pseudo-rapidity between the charged particles in the TPC and those in 
the VZERO detectors greatly suppresses correlations unrelated to the 
azimuthal asymmetry in the initial geometry (``non-flow''). Note that 
the contribution from flow fluctuations was shown to be positive for 
$v_2\{\rm{EP}\}$~\cite{Voloshin:2008dg}.

\begin{figure}[tp]
  \centering
  \includegraphics[keepaspectratio, width=0.3275\columnwidth]{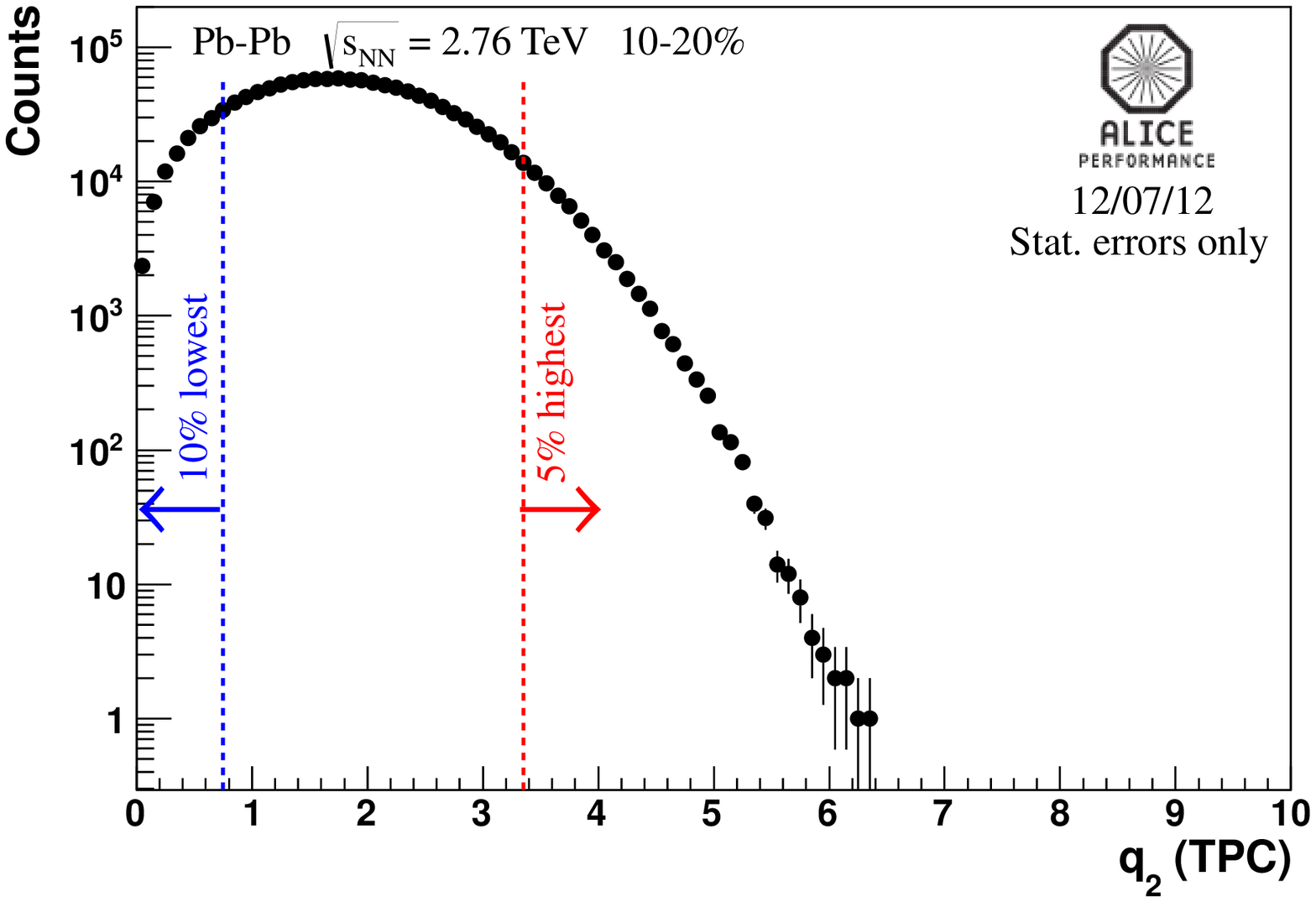}
  \includegraphics[keepaspectratio, width=0.3275\columnwidth]{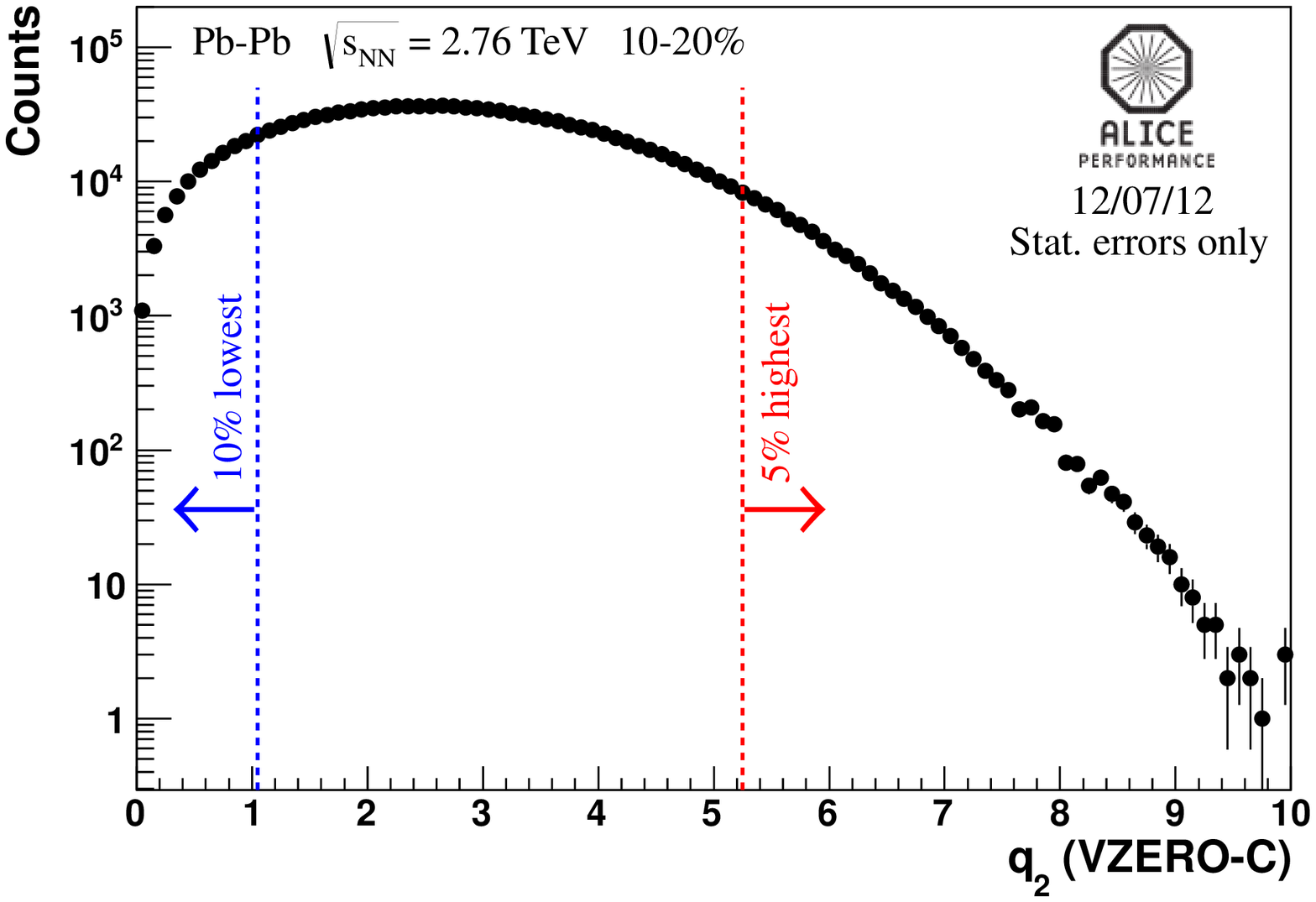}
  \includegraphics[keepaspectratio, width=0.3275\columnwidth]{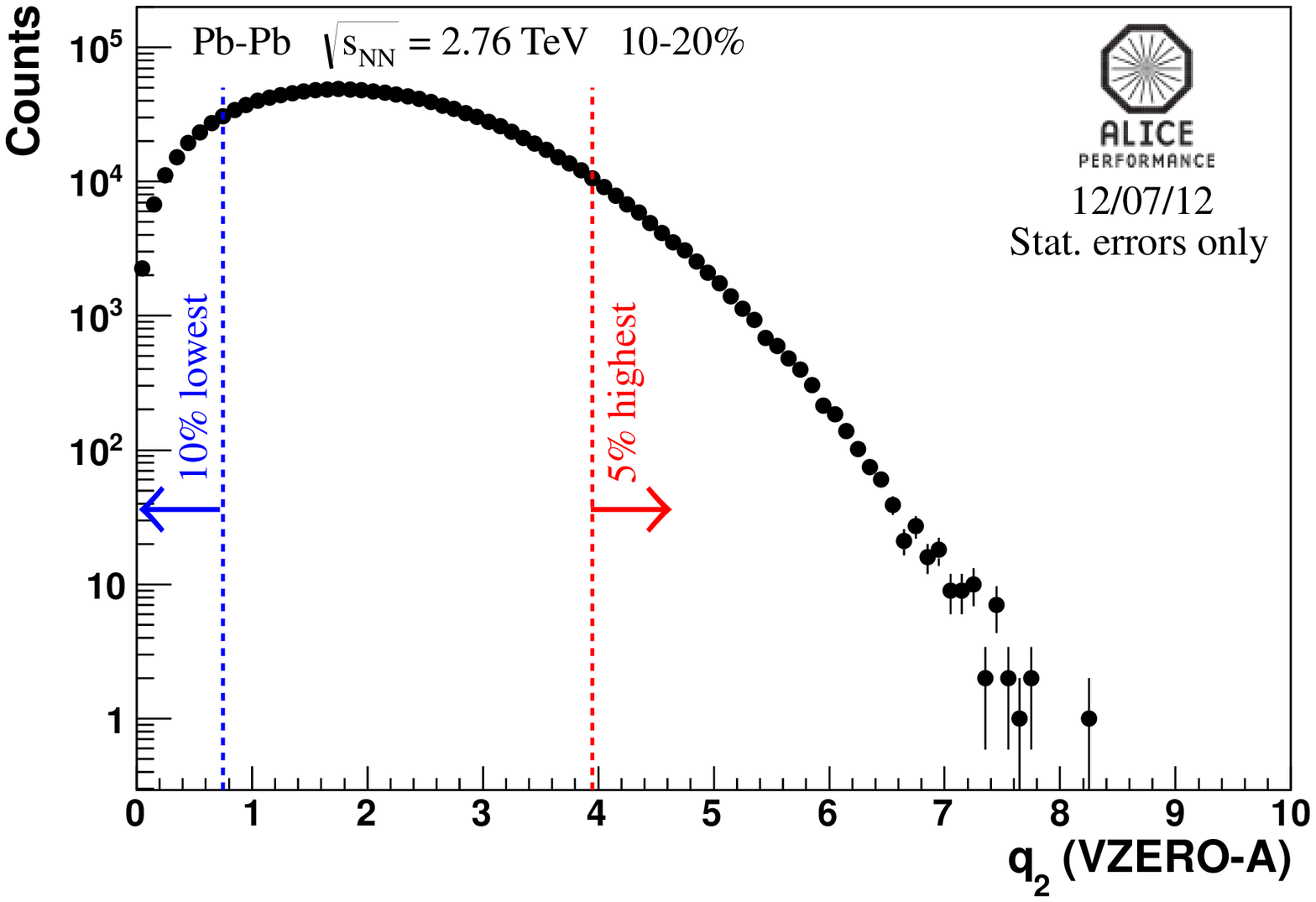}
  \caption{(color online) Distribution of $q_2$ from the TPC (left), 
    VZERO-C (middle), and VZERO-A (right) for the 10-20\% centrality 
    class. The dashed lines indicate the $q_2$ value used to 
    select events with 10\% lowest (blue) and 5\% highest (red) $q_2$, 
    respectively. Only statistical errors are shown.}
  \label{fig:q_dist}
\end{figure}

The $q_2$ distribution is determined in one $\eta$ window of the TPC 
($-0.8<\eta<0$ or $0<\eta<0.8$) as well as in each of the two VZERO 
detectors, see Fig.~\ref{fig:q_dist}. To demonstrate the ability of 
the event shape selection, we compared $v_2$ measured for two classes 
of events: one selected based on events with 10\% lowest and another 
with 5\% highest values of $q_2$ defined for subevent ``a''. There are 
two main effects which define the performance and systematics of the 
event shape selection: kinematic (e.g. pseudo-rapidity) coverage of a 
given detector, and non-flow correlations between subevents involved 
in event selection and $v_2$ calculations for selected class of 
events. The non-flow contributions can be controlled (suppressed) by 
choosing detectors (subevents) which have large rapidity separation 
between each other.

\begin{figure}[tp]
  \centering
  \includegraphics[keepaspectratio, width=0.44\columnwidth]{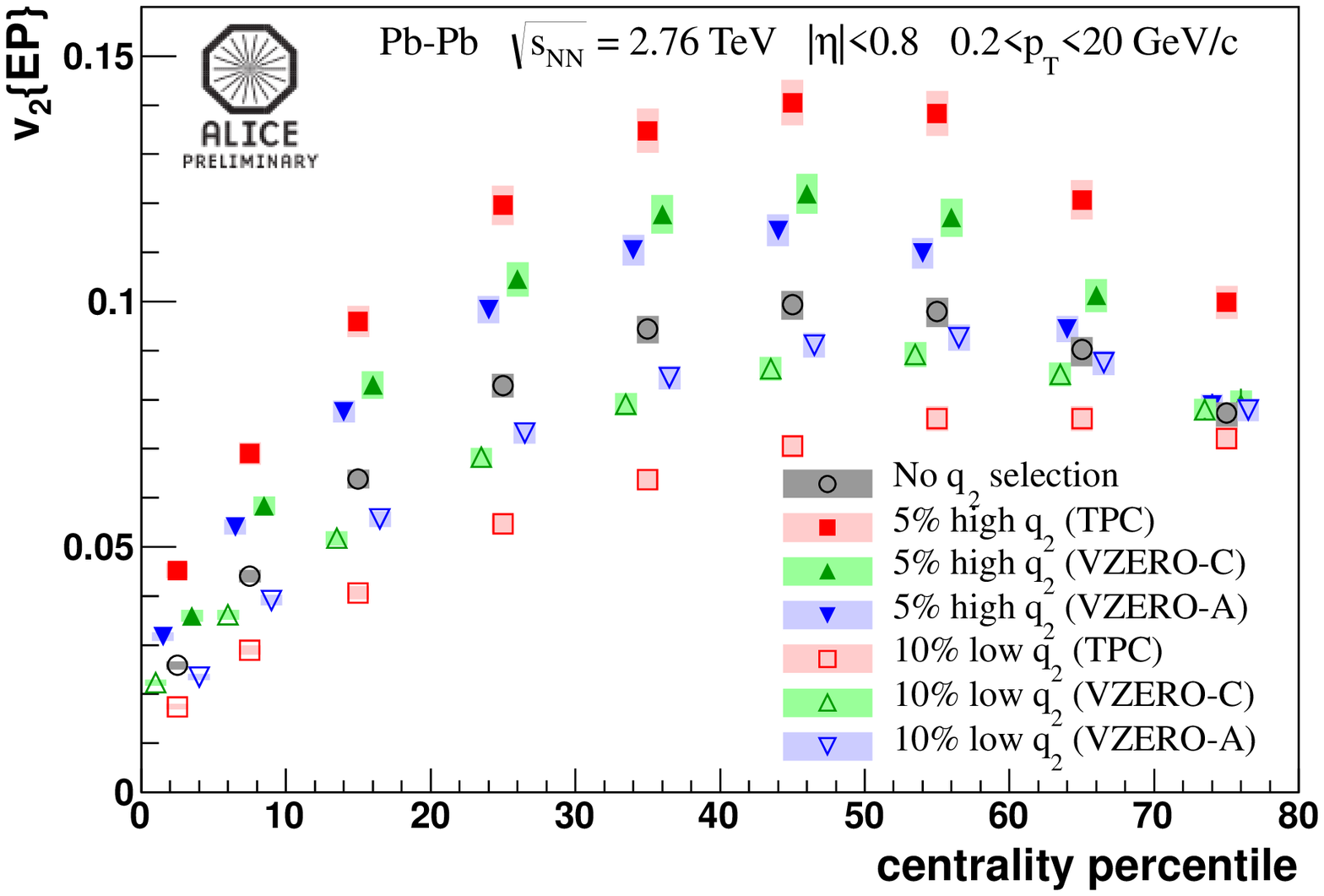}
  \includegraphics[keepaspectratio, width=0.44\columnwidth]{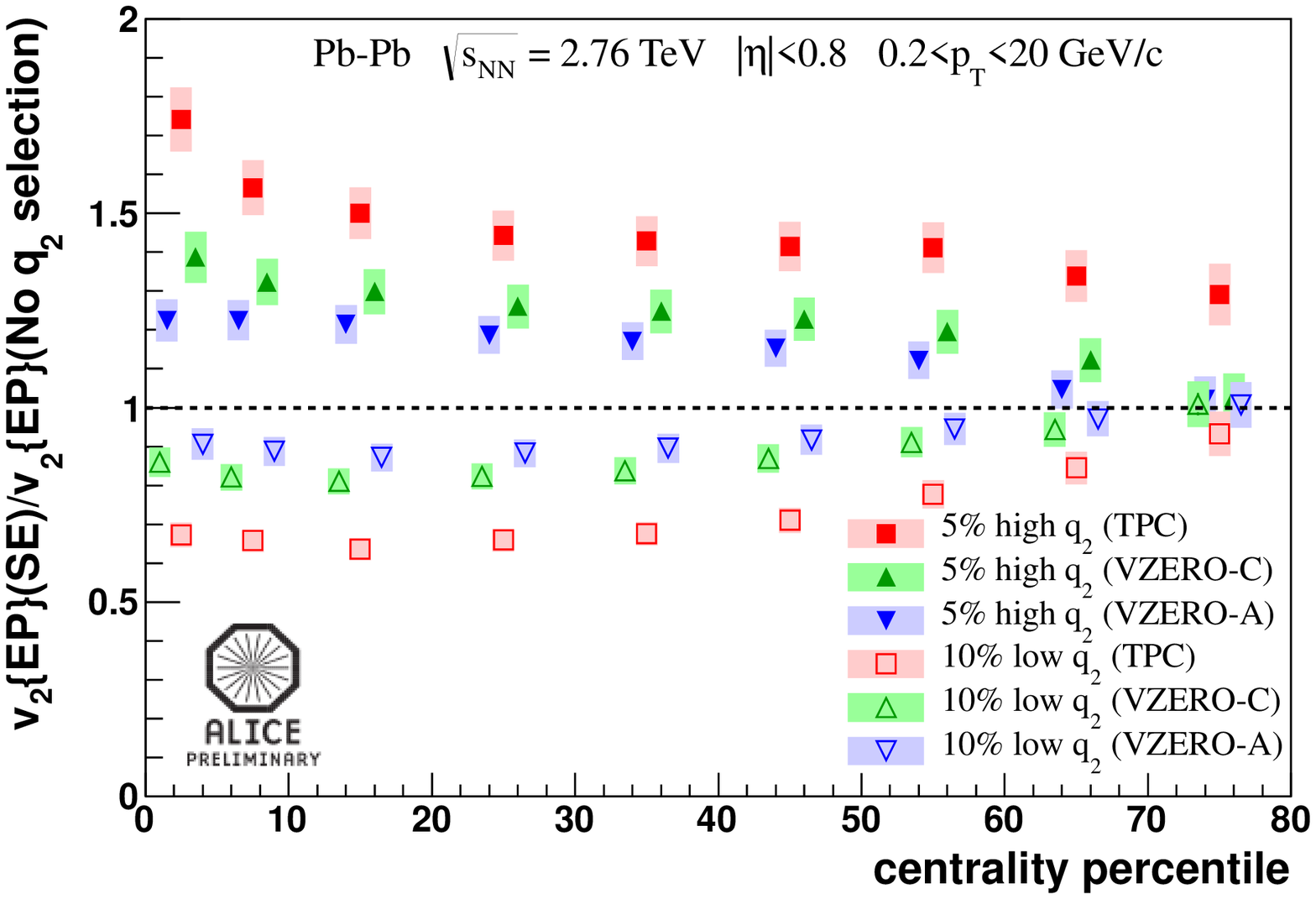}
  \caption{(color online) Unidentied charged particle $v_2$ for event 
    shape selected and unbiased events (left) and their ratios (right) 
    as a function of collision centrality. Central (peripheral) 
    collisions correspond to small (large) values of the centrality 
    percentile. The event selection is based on $q_2$ determined in 
    TPC, VZERO-C, and VZERO-A. For clarity, the markers for shape 
    engineered results (selection based on $q_2$ determined in VZERO-C 
    and VZERO-A) are slightly shifted along the horizontal axis. Error 
    bars (shaded boxes) represent the statistical (systematic) 
    uncertainties. The unbiased results are taken from~\cite{Abelev:2012di}.}
  \label{fig:v2_int}
\end{figure}

\section{Results}

\begin{figure}[bp]
  \centering
  \includegraphics[keepaspectratio, width=0.44\columnwidth]{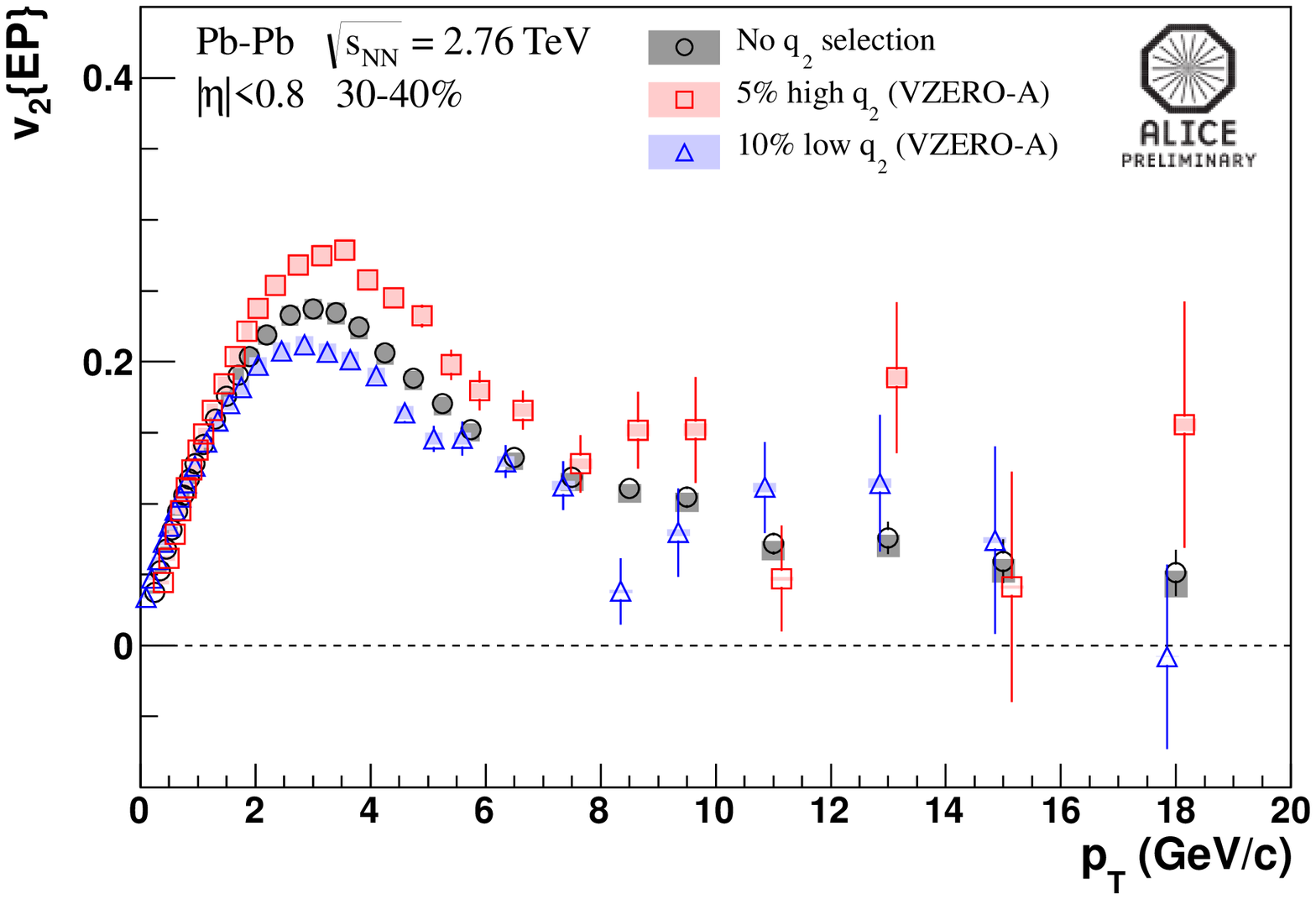}
  \includegraphics[keepaspectratio, width=0.44\columnwidth]{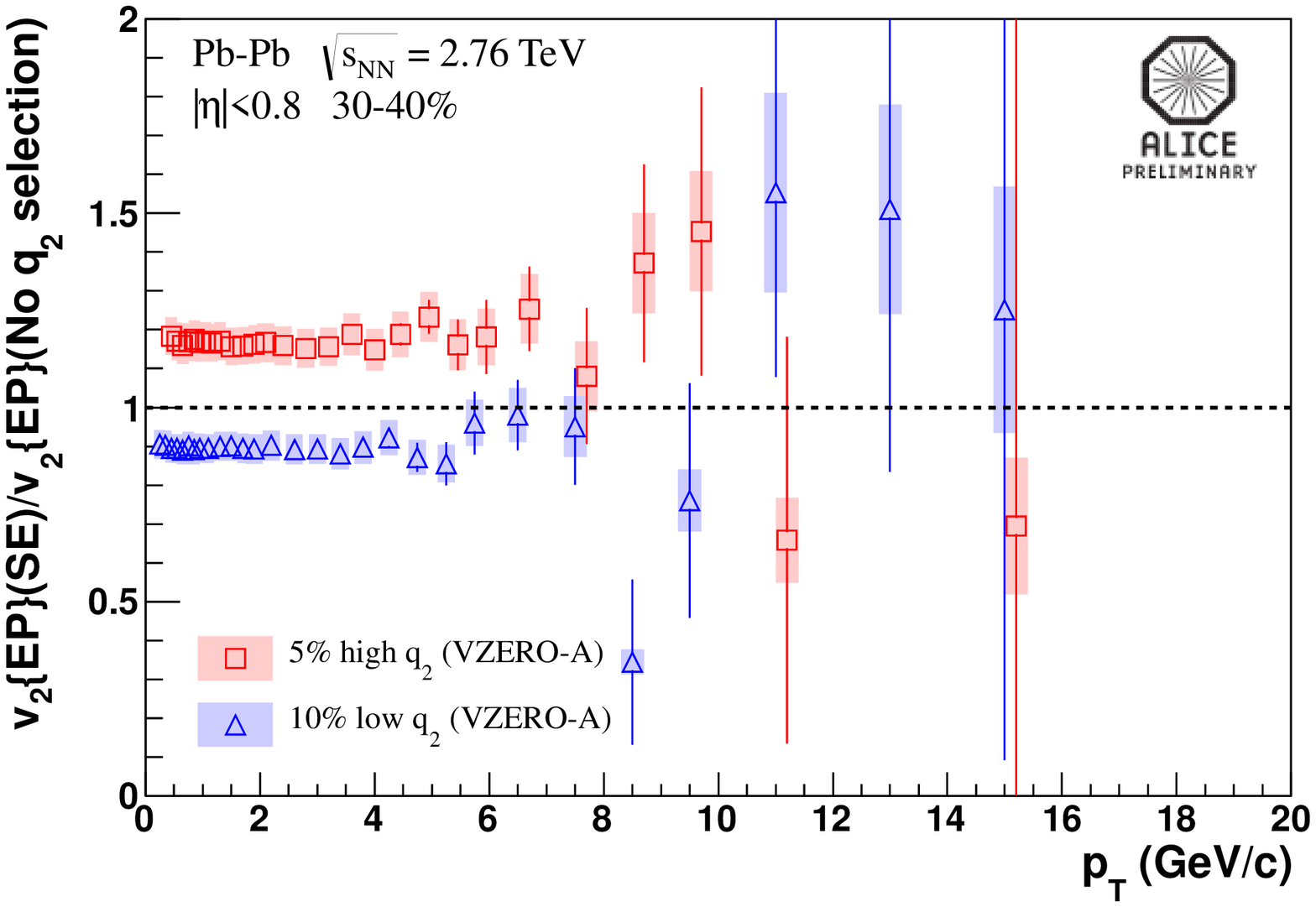}
  \caption{(color online) $v_2(\pt)$ of unidentied charged 
    particles for event shape selected and unbiased events (left) and 
    their ratios (right) for 30-40\% centrality class. The event 
    selection is based on $q_2$ determined in VZERO-A. For clarity, 
    the markers for shape engineered results are slightly shifted 
    along the horizontal axis. Error bars (shaded boxes) represent 
    the statistical (systematic) uncertainties. The unbiased 
    results are taken from~\cite{Abelev:2012di}.}
  \label{fig:v2_pt}
\end{figure}

Figure~\ref{fig:v2_int} shows the unidentified charged particle $v_2$ 
averaged over $0.2<\pt<20$~GeV/$c$ as a function of collision 
centrality for event shape selected and unbiased samples. $v_2$ for 
events with the 5\% highest (10\% lowest) $q_2$ values is larger (smaller) 
than that for events without $q_2$ selection. Results are consistent 
for event selection based on $q_2$ determined in VZERO-C and VZERO-A, 
while results with $q_2$ from TPC differ significantly mainly due to large 
non-flow contributions. Sensitivity of the event shape selection 
deteriorates for peripheral collisions due to small multiplicity and 
reduced magnitude of flow. Only results obtained with the event selection 
using VZERO-A detector which yields the strongest reduction of non-flow 
are reported next.

Figure~\ref{fig:v2_pt} (left) demonstrates that the unidentified 
charged particle $\pt$-differential elliptic flow, $v_{2}(\pt)$, 
differs for event shape selected and unbiased events. The flatness of 
the ratio between $v_2(\pt)$ for event shape selected and unbiased 
events indicates that flow fluctuations are similar at least up to 
$\pt=6$ GeV/$c$ independent of the magnitude of 
the initial anisotropy of the event. For $\pt>6$ GeV/$c$, the effect 
of flow fluctuations may become small though currently large 
experimental uncertainties does not allow to make a firm conclusion.

\begin{figure}[tp]
  \begin{center}
  \includegraphics[keepaspectratio, width=0.44\columnwidth]{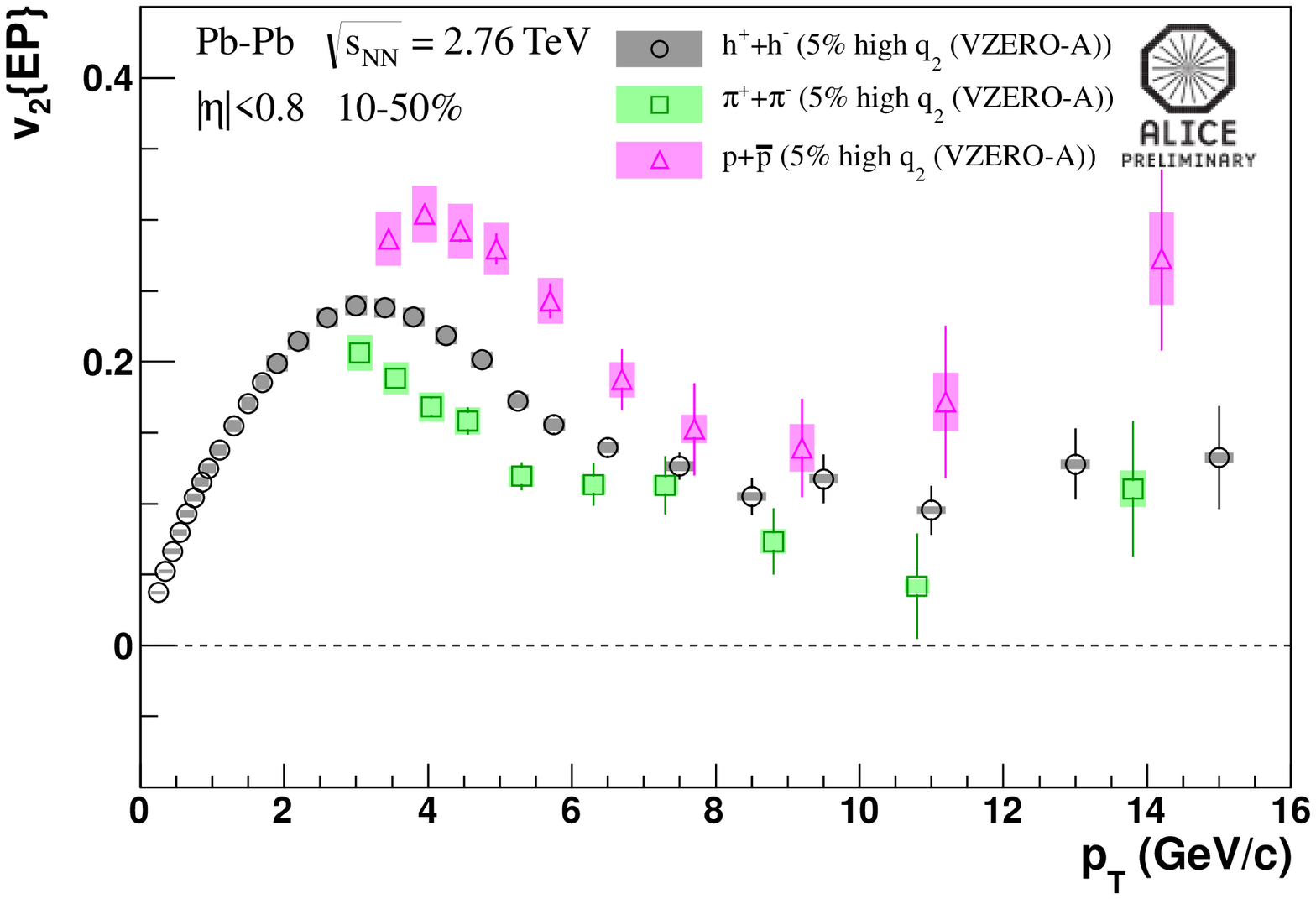}
  \includegraphics[keepaspectratio, width=0.44\columnwidth]{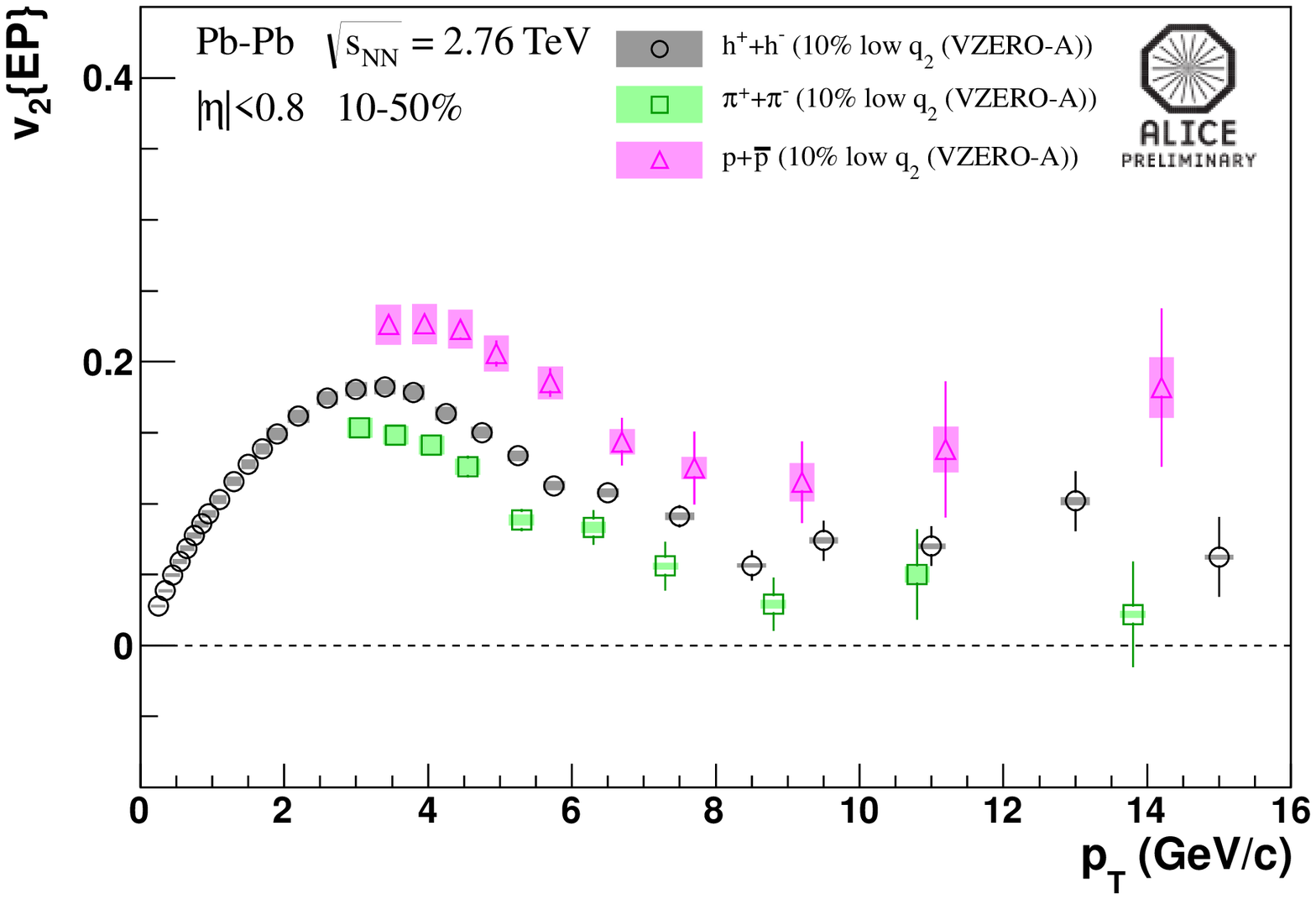}
  \caption{(color online) $v_2(\pt)$ of charged pion and proton for 
    event shape selected events compared to unidentified charged 
    particle results for 10-50\% centrality range. The event selection 
    is based on $q_2$ determined in VZERO-A. For clarity, the markers 
    for pion and proton $v_2$ are slightly shifted along the 
    horizontal axis. Error bars (shaded boxes) represent the 
    statistical (systematic) uncertainties.}
  \label{fig:v2_pid}
  \end{center}
\end{figure}

We also studied the effect of the event shape selection on charged 
pion and proton $v_2$ in comparison to unidentified charged particle 
$v_2$. Figure~\ref{fig:v2_pid} shows this comparison as a function of 
transverse momentum in the 10-50\% centrality range for event shape 
selected events. The proton $v_2$ is higher than that of pions out to 
$\pt=8$ GeV/$c$ where the uncertainties become large, which is similar 
to the result for the unbiased sample~\cite{Abelev:2012di}.

\section{Summary}

We demonstrated that event shape selection based on the azimuthal 
asymmetry of the event can be used to select event samples with 
elliptic flow significantly larger or smaller than the average. This 
opens many new possibilities to study the properties of the system 
created in high energy nucleus-nucleus collisions.

\end{document}